\documentclass[twocolumn,9pt]{article} 

\usepackage[square,numbers,sort&compress,comma]{natbib}

\usepackage{amsmath}
\usepackage{amssymb}
\usepackage{caption}
\usepackage{graphicx}
\usepackage{latexsym}
\usepackage{times}
\usepackage{bm}
\usepackage{upgreek}
\usepackage[pagewise]{lineno}
\usepackage[hidelinks,driverfallback=dvipdfm]{hyperref}
\hypersetup{
colorlinks = true,	
urlcolor = blue,	
linkcolor= blue,	
citecolor = blue 
}
\usepackage[version=4]{mhchem}
\usepackage{tabularx}
\usepackage{multirow}
\usepackage{array}
\usepackage{siunitx}
\usepackage{lipsum}
\usepackage{comment}
\usepackage{xcolor}

\topmargin - 12pt 
\renewenvironment{abstract}%
              {
               \small
               {\bfseries \abstractname}
               \par
               \vspace{10pt}
              }

\renewcommand\abstractname{Abstract}

\newcommand{\nomenclature}
              [1]
              {
               \bgroup
               \flushleft
               \small\bf
               #1
               \par
               \egroup
              }

\renewcommand{\section}
              [1]
              {
               \bgroup
               \flushleft
               \small\bf
               \refstepcounter{section}
               \arabic{section}. #1
               \par
               \egroup
              }

\renewcommand{\subsection}
              [1]
              {
               \bgroup
               \flushleft
               \small\em
               \refstepcounter{subsection}
               \arabic{section}.
               \arabic{subsection}. #1
               \par
               \egroup
              }

\renewcommand{\subsubsection}
              [1]
              {
               \bgroup
               \flushleft
               \small\em
               \refstepcounter{subsubsection}
               \arabic{section}.
               \arabic{subsection}.
               \arabic{subsubsection}. #1
               \par
               \egroup
              }

  \newcommand{\acknowledgement}
              [1]
              {
               \bgroup
               \flushleft
               \small\bf
               #1
               \par
               \egroup
              }

  \newcommand{\sectionbib}
              [1]
              {
               \bgroup
               \flushleft
               \small\bf
               #1
               \par
               \egroup
              }

\newcommand\T{\rule{0pt}{2.6ex}}       
\newcommand\B{\rule[-1.2ex]{0pt}{0pt}} 

\newcommand{\rt}[1]{\mathrm{#1}} 
\newcommand{\bmss}[1]{\bm{\mathsf{#1}}}
\usepackage[pdftex]{pdfpages} 
\usepackage{pgffor} 

\makeatletter
\AtBeginDocument{\let\LS@rot\@undefined}
\makeatother

\def\supplementfilename{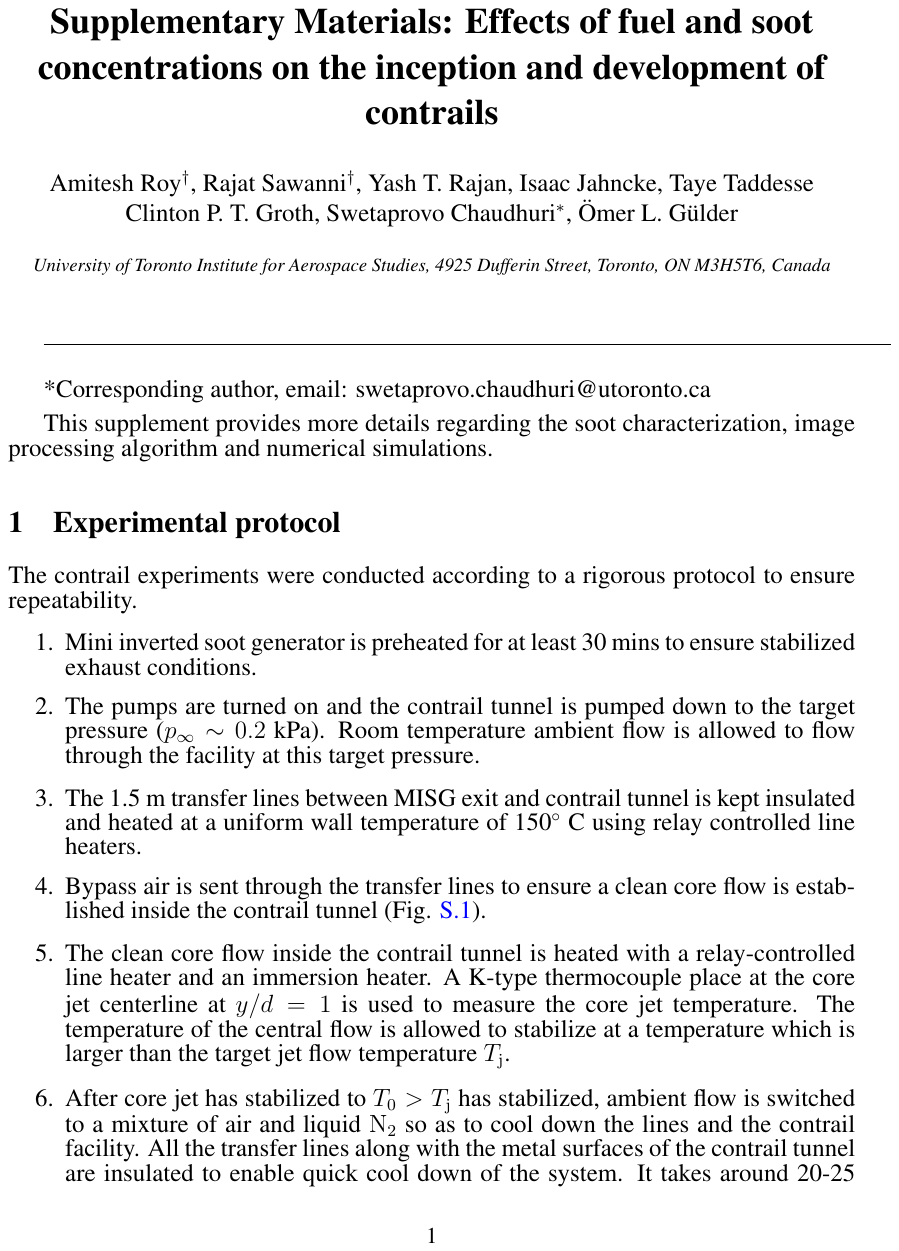}

\pdfximage{\supplementfilename}
\def\numbersupplementpages{\the\pdflastximagepages}

\newif\ifarXiv
\arXivtrue 

\begin{document}



\small
\baselineskip 10pt

\setcounter{page}{1}

\title{\LARGE \bf Effects of fuel and soot concentrations on the inception and development of contrails}

\author{{\large Amitesh Roy$^{\dagger}$, Rajat Sawanni$^{\dagger}$, Yash T. Rajan,  Isaac Jahncke, Taye Taddesse}\\ 
        {\large Clinton P. T. Groth, Swetaprovo Chaudhuri$^{*}$, \"{O}mer L. G\"{u}lder}\\[10pt]
        {\footnotesize \em University of Toronto Institute for Aerospace Studies, 4925 Dufferin Street, Toronto, ON M3H5T6, Canada}}

\date{}  

\twocolumn[\begin{@twocolumnfalse}
\maketitle
\rule{\textwidth}{0.5pt}
\vspace{-5pt}

\begin{abstract} 
\vspace{-5pt}
 Fundamental questions related to the roles of fuel type, combustion parameters, and turbulence transport interactions in the inception and growth of contrails have remained intractable in remote sensing and in-flight measurements. Consequently, we developed a novel laboratory-scale facility for studying the inception, growth and persistence of contrails for aircraft-relevant conditions. The set of exhaust conditions, generated using an inverted co-flow soot generator at a set of global equivalence ratio for two fuels - ethylene and propane, is supplied to the contrail tunnel which then mixes with an ambient flow emulating long-haul aircraft cruise conditions (\SI{20.8}{kPa} and \SI{190}{K}). Detailed soot characterization using a scanning mobility particle sizer and transmission electron microscopy is coupled with measurements of instantaneous and averaged scattering intensities from the generated contrails. The experimental results are complemented by numerical simulations of the contrail tunnel using solutions of the Favre-averaged Navier-Stokes (FANS) equation and a two-equation model for handling particulate matter, including soot and ice. Results show the first experimental snapshots of a contrail cross section, highlighting the interaction of turbulent mixing and microphysical growth scales involved in ice nucleation across the shear layers. As expected, the average scattering intensities of contrails increase with soot number concentrations and water vapor content. Comparisons between ethylene and propane exhausts indicate that the scattering propensity of contrails is more sensitive to exhaust water vapor content than to soot concentrations. Finally, depolarization measurements are used to show asphericity in ice crystal habits. Thus, our study present a unique window into contrail formation, theoretical modeling and simulation.
\end{abstract}

\vspace{5pt}

{\bf Novelty and significance statement}

\vspace{5pt}
The role of hydrocarbon combustion exhaust and soot concentrations on contrail formation propensity is investigated in a novel contrail tunnel that emulates the turbofan exit conditions of an aircraft cruising at upper troposphere. Sensitivity of ice scattering is explored in unprecedented details through experiments, modeling and numerical simulations, revealing complex interplay of turbulence, mixing, microphysics and ice nucleation.

\vspace{5pt}
\parbox{1.0\textwidth}{\footnotesize {\em Keywords:} Contrails; Soot; Ice nucleation; Mie Scattering; Depolarization}
\rule{\textwidth}{0.5pt}
$^\dagger$ Contributed equally. *Corresponding author, email: swetaprovo.chaudhuri@utoronto.ca 
\vspace{5pt}
\end{@twocolumnfalse}] 

\section{Introduction\label{sec:introduction}} \addvspace{10pt}
\vspace{-5pt}
Aircraft-induced cloudiness (AIC) is the largest contributor to effective radiative forcing from aviation-derived radiative forcing (RF) \cite{lee2021contribution}, surpassing combined forcings from CO\textsubscript{2} and NO\textsubscript{x}. Due to its short-lived nature, AIC is prone to rapid mitigation strategies \cite{martin2024feasibility, engberg2025forecasting}. However, the poor level of scientific understanding has proved to be a major impediment in the development of global mitigation strategies \cite{IPCC2021}.

\begin{figure*}[h!]
	\centering
	\includegraphics[width=\textwidth]{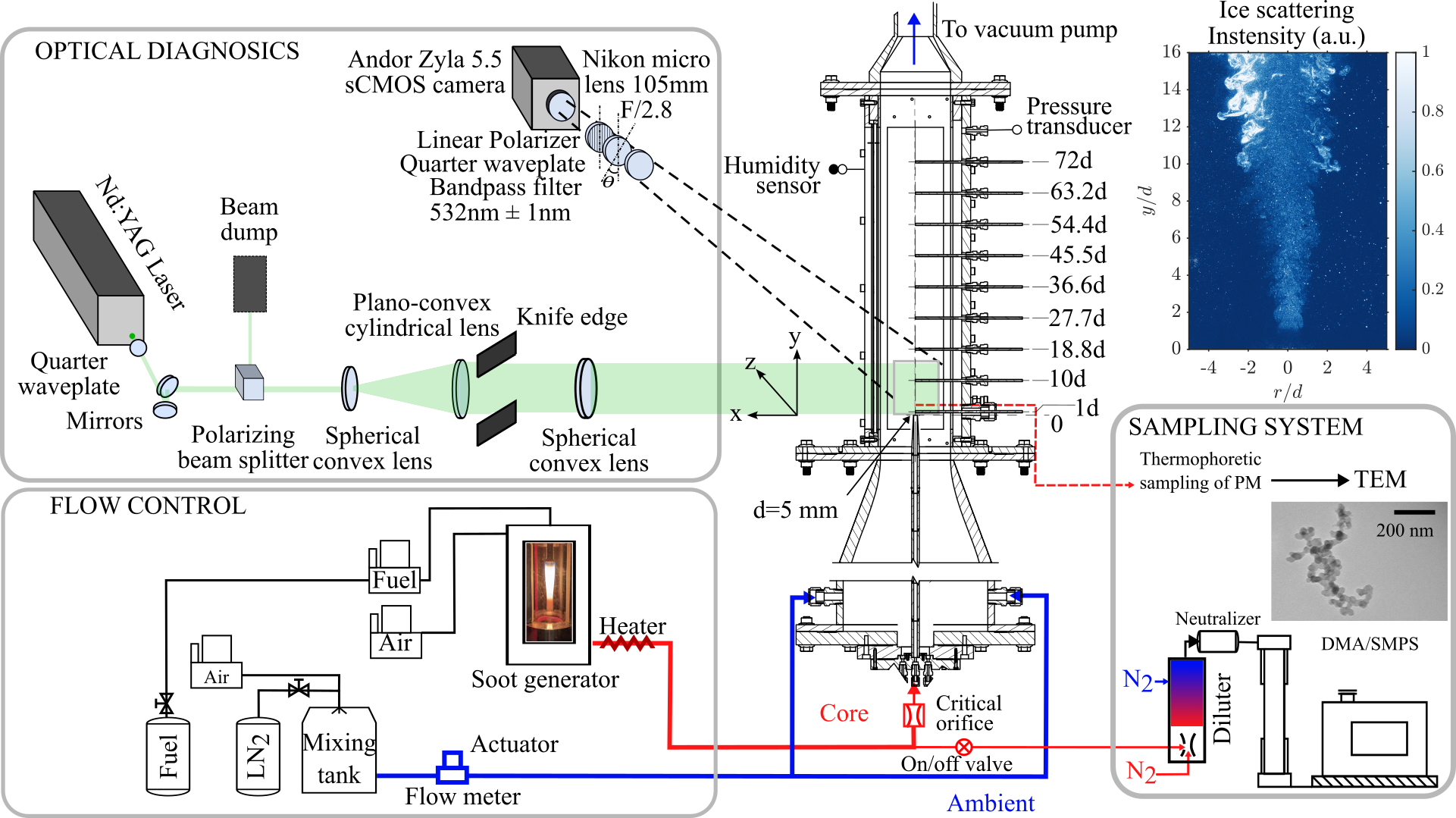}
	\caption{\footnotesize Schematic of the contrail facility developed at the University of Toronto. Aircraft engine exit conditions at cruising altitude is simulated by the ambient and core jet flow in the test section where various measurements enable contrail quantification. (Right) Snapshot of instantaneous scattering due to ice formation on exhaust generated from $\phi=0.161$ \ce{C2H4} flame.}
	\label{fig:PID}
    \vspace*{-15pt}
\end{figure*}

Contrails, a subset of AIC and precursors to persistent cirrus clouds, form when the hot, moist, and particulate-laden combustion exhaust mixes with the cold, humid air near the low-pressure upper troposphere - lower stratosphere (UTLS, \SI{8}{km}-\SI{14}{km}) altitudes, triggering condensation and freezing of water vapor into cloud trails. The threshold condition for contrail formation and its persistence consider isobaric mixing from exhaust gas to ambient surroundings, and is referred to as the Schmidt-Appleman criteria (SAC) \cite{schmidt1941entstehung, appleman1953formation}. SAC identifies the threshold condition as the point where exhaust dilution causes a proportionate decrease in water vapor and plume temperature (represented using a straight mixing line in $p_\rt{vap}-T$ diagram) to cross the liquid saturation curve, while its persistence is ensured by ambient supersaturation with respect to ice. For prediction of observable properties of contrails such as the size, concentration and optical depth of ice particles, SAC is complemented by aerosol microphysics and flow modeling  \cite{karcher2015microphysical, ponsonby2025updated, paoli2004contrail, paoli2016contrail, khou2015spatial}, and validated against remote-sensing \cite{spichtinger2003ice, agarwal2022reanalysis} and in-flight measurement data \cite{voigt2017ml, brauer2021airborne, moore2017biofuel}. 

Thermodynamics ($T$) and the water vapor composition of the exhaust ($x_{\ce{H2O}}^\rt{v}$) and ambient streams control the slope of the mixing line, while the turbulence and momentum ratios of the two streams \cite{charonko2017variable, pitts1991effects} control the cooling rate \cite{karcher1994transport, paoli2016contrail, fries2023lagrangian, barahona2012ice}. Depending on the supersaturation, cooling rates, and physicochemical properties of the particulate matter (PM), various modes of heterogeneous and homogeneous
ice nucleation \cite{hoose2012heterogeneous, knopf2023atmospheric} can be triggered during the contrail inception process. Among the particulates, the non volatile soot particles are initially hydrophobic \cite{seisel2005water}, but can become hydrophilic after being coated with volatile exhaust constituents \cite{dischl2025fuel, gao2022dependence, gao2022laboratory, schnitzler2017coating} and through atmospheric aging \cite{zhang2008variability, mahrt2020aging, testa2024simulated}; the concentrations of such coated and aged soot have been found to correlate with concentrations of ice particles in highly sooting exhaust plumes \cite{karcher2018formation}. However, recently, jet engines running with cleaner sustainable aviation fuels have shown that contrails might remain prominent even under low sooting conditions \cite{moore2017biofuel, voigt2025substantial}, highlighting the roles of volatiles, aqueous aerosols, lubrication oil and ambient aerosols.

Due to its multi-physical nature, modeling of contrails has remained a significant challenge which motivates direct \textit{in-situ} measurements under contrail‑relevant conditions. Field campaigns \cite{voigt2025substantial, brauer2021airborne, moore2017biofuel} that could provide such data are expensive and rarely yield resolved measurements with well characterized boundary conditions. \textit{Ex-situ} laboratory studies using diffusional ice nucleation chambers \cite{testa2024simulated, gao2022laboratory, gao2022enhanced} offer sensitive control necessary to quantify specific nucleation modes, but cannot reproduce the full mixing‑line trajectory or the rapid cooling rates typical of turbulent shear layers, limiting their applicability under contrail-relevant conditions. Lab-scale facilities \cite{fagan2024contrails, wong2013laboratory} offer a unique middle ground, but have not been realized to their full potential to produce a more complete characterization of contrails through resolved diagnostic capabilities. 


The present study addresses these gaps by providing \textit{in-situ} quantification of ice formation in a newly developed contrail-in-a-lab facility. We combine detailed soot characterization and two-dimensional Mie-scattering imaging coupled with Stokes polarimetry \cite{schaefer2007measuring, azzam2016stokes} to show, for the first time, how exhaust containing differing concentrations of soot and water vapor, generated from two fuels -- ethylene and propane -- affect the structure of contrail formation within the turbulent mixing layer of hot exhaust and cold ambient. Experimental observations are complemented by theoretical predictions and simulation of the Favre-averaged Navier-Stokes (FANS) equation coupled with a semi-empirical two-equation model for the dispersed phase \cite{chelem2025contrail,jahncke2026fans, khou2015spatial}. With well characterized boundary conditions, numerical predictions from the simplistic diffusional ice growth model can be translated and compared to experimental scattering intensities using the Mie theory \cite{prahl2024miepython}. Finally, depolarization measurements are used to show the aspheric nature of the nucleated ice particles. Thus, our study provides experimentally constrained benchmarks for interpreting contrail variability and to improve microphysical parameterizations under contrail‑relevant conditions.


\section{Methodology\label{sec:methods}} \vspace{-5pt}
\subsection{Test rig and flow conditions} \addvspace{5pt}

We perform experiments in a newly developed wind tunnel facility designed specifically for recreating thermodynamic and flow conditions encountered at the engine exhaust of cruising aircrafts at UTLS altitudes. The facility, shown in Fig. \ref{fig:PID}, consists of various control systems that maintain a low-pressure cold ambient flow and a central hot jet of exhaust produced by a soot generator. The primary test section consists of a $\SI{0.5}{m}$ long tunnel with a square cross section of $\SI{0.1}{m}$ width. Three optically accessible cast acrylic windows that span the length of the tunnel provide a unique view of the contrail formation process, while the fourth side houses instrumentation, allowing for controlled monitoring of pressure, temperature, and exhaust properties along the length of the tunnel.

During the run, the test section is maintained at a reference temperature of $T_{\mathrm{ref}}=190\pm2$ K and a pressure of $p_\infty=20.8\pm 0.15$ kPa to emulate a cruising altitude of 12.2 km. Pressure is maintained using two vacuum pumps downstream of the tunnel, while the cold ambient co-flow is generated by mixing controlled amounts of liquid and gaseous \ce{N2} to regulate ambient humidity. The ambient saturation at the exit of the test section was measured as $s^\rt{v}_\infty=0.35$ (Rotronic HC2A humidity probe) at $T=\SI{210}{K}$. The ambient flow is conditioned using honeycombs and beads before entering the test section through a contraction with an area ratio of 3.8. The nominal flow velocity is maintained at $\bar{v}_\infty=5.6$ m/s, ensuring turbulent conditions ($Re_{\infty} = 14.4\times10^4$).

The heated exhaust jet is emitted from an insulated straight tube with $d=\SI{5}{mm}$ and a wall thickness of 3 mm following a slow taper at the exit plane. The jet exhaust is maintained at an exit temperature of $T_{\mathrm{j}}= 560$ K and a mean velocity of $\bar{v}_\rt{j}=83.25$ m/s, ensuring near turbulent conditions $Re_{\mathrm{j}} = 2.18\times10^3$. Temperature-controlled choked orifices are utilized to control the flow rate of unfiltered exhaust using heated feed lines between the soot generator and the central nozzle. Exhaust is generated using the Argonaut mini-inverted soot generator (MISG) comprising an inverted, laminar co-flow diffusion flame that burns gaseous fuel in over-ventilated conditions \citep{kazemimanesh2019novel, moallemi2019characterization}. Although not representative of aircraft soot, the Argonaut burner was chosen here to ensure well-controlled and reproducible soot properties without introducing additional parametric effects. We use ethylene (\ce{C2H4}) and propane (\ce{C3H8}) as fuel at three global equivalence ratios: $\phi = 0.186, 0.161, 0.124$ with fixed mass flow of air of 10 slpm to generate a controlled range of soot concentration ($N_{\mathrm{s}}$) and water vapor mass fraction ($x^\rt{v}_{\ce{H2O}}$). To keep the parameter space manageable and avoid the introduction of confounding effects, the thermodynamic and flow conditions are kept constant throughout all experiments, and only the core jet exhaust composition is varied (see Table \ref{tab:exp_cond}). A rigorous experimental protocol was developed to ensure close control of flow variables and ensure measurement repeatability and is described in more detail in \S1 of Supplementary Information (SI).

\begin{table}[t]
    \centering
    \caption{\footnotesize Test conditions. Notation - $\phi$: Global equivalence ratio, $N_{\mathrm{s}}$: total soot number density, $\bar{d}_{\mathrm{m}}$: Mean mobility diameter, $\mathcal{D_\mathrm{F}}$: fractal dimension, $k_\mathrm{g}$: fractal pre-factor, $\bar{d_\mathrm{p}}$: Mean primary particle diameter. $\sigma_{(\cdot)}$ indicate the standard deviations. See SI \S2.2 for error quantification.}
    \resizebox{\columnwidth}{!}{
    \begin{tabular}{|c|c|c|c|c|c|c|c|c|c|c|}
        \hline
Fuel & $\phi$ & $x^\rt{v}_{\ce{H2O}}$ 
& \multicolumn{2}{c|}{$N_{\mathrm{s}}$, $\sigma_{N_{\mathrm{s}}}$} 
& \multicolumn{2}{c|}{$\bar{d}_\mathrm{m}$, $\sigma_{\bar{d}_\mathrm{m}}$} 
& \multicolumn{2}{c|}{$\mathcal{D}_\mathrm{F}$, $\sigma_{\mathcal{D}_\mathrm{F}}$} 
& $k_\mathrm{g}$ 
& $d_\mathrm{p}$ \B\\

& & 
& \multicolumn{2}{c|}{$10^{12}\ \mathrm{m}^{-3}$} 
& \multicolumn{2}{c|}{nm} 
& \multicolumn{2}{c|}{} 
&  
& nm \T\B\\
        \hline
        \ce{C2H4} & 0.124 & 0.0172 & 7.13 & 0.41 & 168.8 & 6.7 & 1.78 & 0.11 & 1.99 & 29.1\T\\
        \ce{C2H4} & 0.161 & 0.0223 & 9.24 & 0.91 & 163.3 & 10.4 & 1.75 & 0.09 & 2.23 & 30.8 \\
        \ce{C2H4} & 0.186 & 0.0257 & 14.2 & 1.75 & 176.1 & 6.3 & 1.59 & 0.06 & 2.88 & 32.5 \\
        \ce{C3H8} & 0.124 & 0.0206 & 0.21 & 0.04 & 123.5 & 19.9 & 1.88 & 0.08 & 1.53 & 33.2 \\
        \ce{C3H8} & 0.161 & 0.0267 & 1.48 & 0.05 & 77.1 & 3.8 & 1.78 & 0.06 & 1.82 & 31.8  \\
        \ce{C3H8} & 0.186 & 0.0308 & 6.74 & 0.63 & 177.5 & 6.7 & 1.70 & 0.07 & 1.96 & 32.3  \\
        \hline
    \end{tabular}}
    \label{tab:exp_cond}
    \vspace*{-10pt}
\end{table}

The temperature field inside the test section was measured using 9 K-type and E-type thermocouples that span the height of the tunnel (see test section in Fig. \ref{fig:PID}). Each thermocouple was moved across the tunnel width to map the radial temperature field at that height during steady-state conditions. The temperature field at any given location in the tunnel is time-averaged and then interpolated between the measured points to obtain Fig. \ref{fig:TempProfile}a.

\subsection{Flow and particulate diagnostics \label{subsec:flow-particle}}\addvspace{5pt}
Particulate measurements employ a Scanning Mobility Particle Sizer (SMPS GRIMM Vienna-L DMA and CPC) and Transmission Electron Microscopy (TEM) after sampling with the contrail tunnel maintained at $T_\infty=\SI{298}{K}$. While TEM images are used to gain an understanding of the fractal properties (mean primary particle diameter, fractal dimension and fractal prefactor), information about the aggregate size and concentration is gained using SMPS measurements of mobility size distribution ($d_\rt{m}$) and $N_\rt{s}$. SMPS sampling is performed in the exhaust lines before entry into the contrail tunnel. The sampling procedure utilizes an internal dual-stage dilution system to achieve a dilution ratio of 232, which is sufficient to measure nvPM (soot) concentrations and mobility size distributions, but ignores vPM contributions \cite{kazemimanesh2019novel, arp6320b2024procedure}. However, the exhaust in the tunnel remains unprocessed and contains both vPM and nvPM. For TEM imaging, soot is collected inside the contrail tunnel at the exit of the central nozzle by briefly exposing carbon-coated \ce{Cu} grids. The grids are then imaged (see the fractal soot aggregate in the inset of Fig. \ref{fig:PID}) in Hitachi HT7700 TEM (100 kV, 0.2 nm resolution). Fractal aggregates are analyzed by calculating the mean diameter of the primary particle ($d_\mathrm{p}$) and the number of primary particles $N_\mathrm{p}$, projected aggregate area ($A_\mathrm{p}$), and longest projected length ($L$) for each aggregate using Fiji, an open-source software \cite{Fiji}. These parameters are then utilized to determine aggregate fractal dimension $\mathcal{D}_\mathrm{F}$ and the pre-factor $k_{\mathrm{f}}$ using a well established methodology \cite{koylu1995langmuir, sipkens2023overview}. Sampling methodology and post-processing is elaborated in SI $\S2$.

\begin{figure}[t]
	\centering
	\includegraphics[width=\columnwidth]{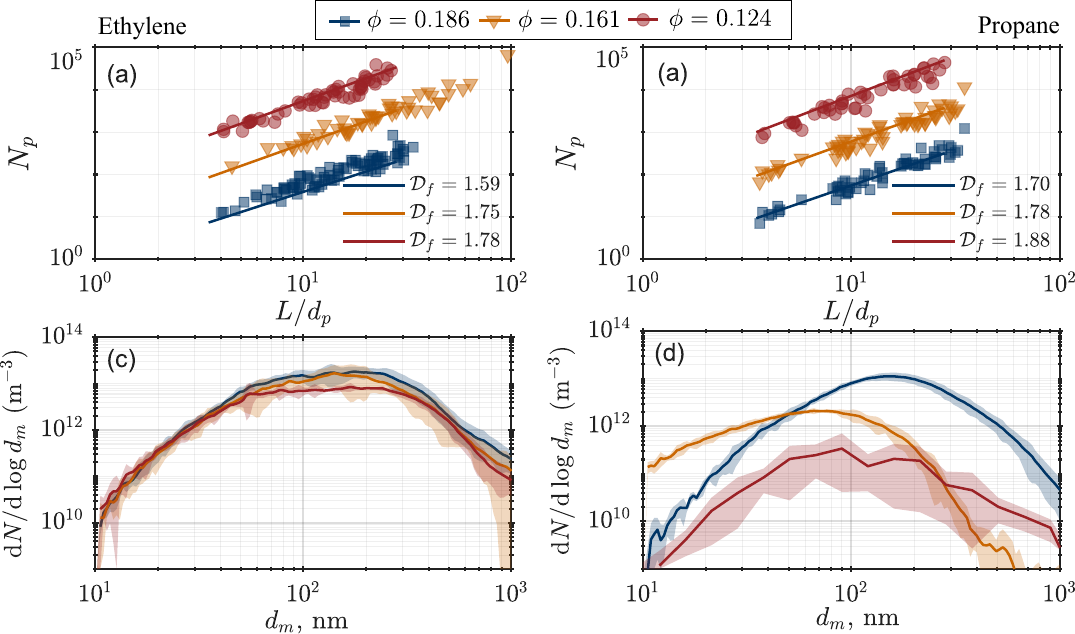}
	\caption{\footnotesize Properties of soot particles generated by the inverse diffusion flame for the two fuel types. (a, b) Observed fractal dimension $\mathcal{D}_\mathrm{F}$ determined from TEM imaging of soot particles. Coefficient of fit $R^2>0.87$ for all cases and plots shifted by a factor of 10 for visual clarity. (c, d) Log-normal soot mobility size distribution obtained through SMPS measurements. See text for error bar details.}
	\label{fig:mobdia}
    \vspace*{-15pt}
\end{figure}

The size distributions $d_\mathrm{p}$ and $d_\mathrm{m}$ are lognormal for all cases, with their specific moments and distributions shown in Table \ref{tab:exp_cond} and Fig. \ref{fig:mobdia}, respectively ($d_\mathrm{p}$ distribution shown in Fig. S3 of SI). The primary particle diameters remain roughly unchanged across all the cases studied. With an increase in $\phi$, a larger number of aggregates $N_{\mathrm{s}}$, with an increasingly open and less compact structure (lower $\mathcal{D}_\mathrm{F}$ and higher $k_\mathrm{g}$ \cite{sipkens2023overview}), are observed for both \ce{C3H8} and \ce{C2H4} flames. For \ce{C3H8} flames, the decrease in $\phi$ is found to cause a much more significant decrease in particle size and number concentrations (see Fig. \ref{fig:mobdia}d), compared to \ce{C2H4} flames. In fact, the mobility diameters for the $\phi=0.161$ \ce{C3H8} flame are noted to be significantly lower than in all other cases. For the lowest sooting case, i.e. those comparable to \ce{C3H8} flame at $\phi=0.124$, mobility diameter distributions and total particle counts have been noted to suffer large fluctuations between days of experiment \cite{moallemi2019characterization}. Consequently, for all cases, the $2\sigma$ error band is shown except for the $\phi=0.124$ \ce{C3H8} case (see Fig. S3 and \S2.2 in SI). For this case, the data is shown after performing moving average over a window size of $2.27$ nm across 21 realizations and the error bar only represents $0.5\sigma$ band. 

Finally, it is worth noting here that MISG exhaust generated from $\ce{C2H4}$ flame have been reported to contain a higher degree ($\geq 90\%$) of elemental carbon to total carbon (EC/TC) than $\ce{C3H8}$ flames ($\geq 70\%$) \cite{kazemimanesh2019novel, moallemi2019characterization, vernocchi2022characterization}. This implies $\ce{C2H2}$ flames produce higher non-volatile PM (nvPM). In comparison, $\ce{C3H8}$ flame produces a higher percentage of organic carbon content leading to higher volatile PM (vPM), which also increases at leaner conditions \cite{vernocchi2022characterization}. This fact is perhaps also reflected in the lower values of $N_\rt{s}$ for $\ce{C3H8}$ flames in Fig. \ref{fig:mobdia}d. The implications of PM composition in the exhaust on ice nucleation is discussed in \S\ref{subsec-meanscattering}.


\subsection{Stokes polarimetry}\addvspace{5pt}
Ice and soot particles in the test section are visualized using a pulsed Nd-YAG laser (Quanta-Ray, \SI{532}{nm}, \SI{300}{mJ/pulse}) at \SI{10}{Hz} using elastic light scattering in 2D. Laser sheet created by beam-forming optics illuminates the central plane between a $y = d$ to $20d$ from the exit plane (see Fig. \ref{fig:PID}). A combination of quarter-wave plate and polarizing beam splitter is used to ensure that the incident laser sheet is perpendicularly polarized with respect to the scattering plane ($x-z$). The scattered light at $90^{\circ}$ is collected using an sCMOS camera (Andor Zyla 5.5 sCMOS) through a Nikon lens (f=105mm and f/2.8) and a 532$\pm$1 nm band-pass filter. The field of view spans 75$\times$97.6 mm imaged onto 1575$\times$2050 pixels at 10 Hz, phase-locked to the laser pulse. Additionally, the polarization state of the scattered light is quantified using a fixed linear polarizer coupled to a quarter-wave plate affixed on a rotating mount. The angle between the two is changed to image different polarization vectors of the scattered intensity. 

\begin{figure}
	\centering
	\includegraphics[width=\columnwidth]{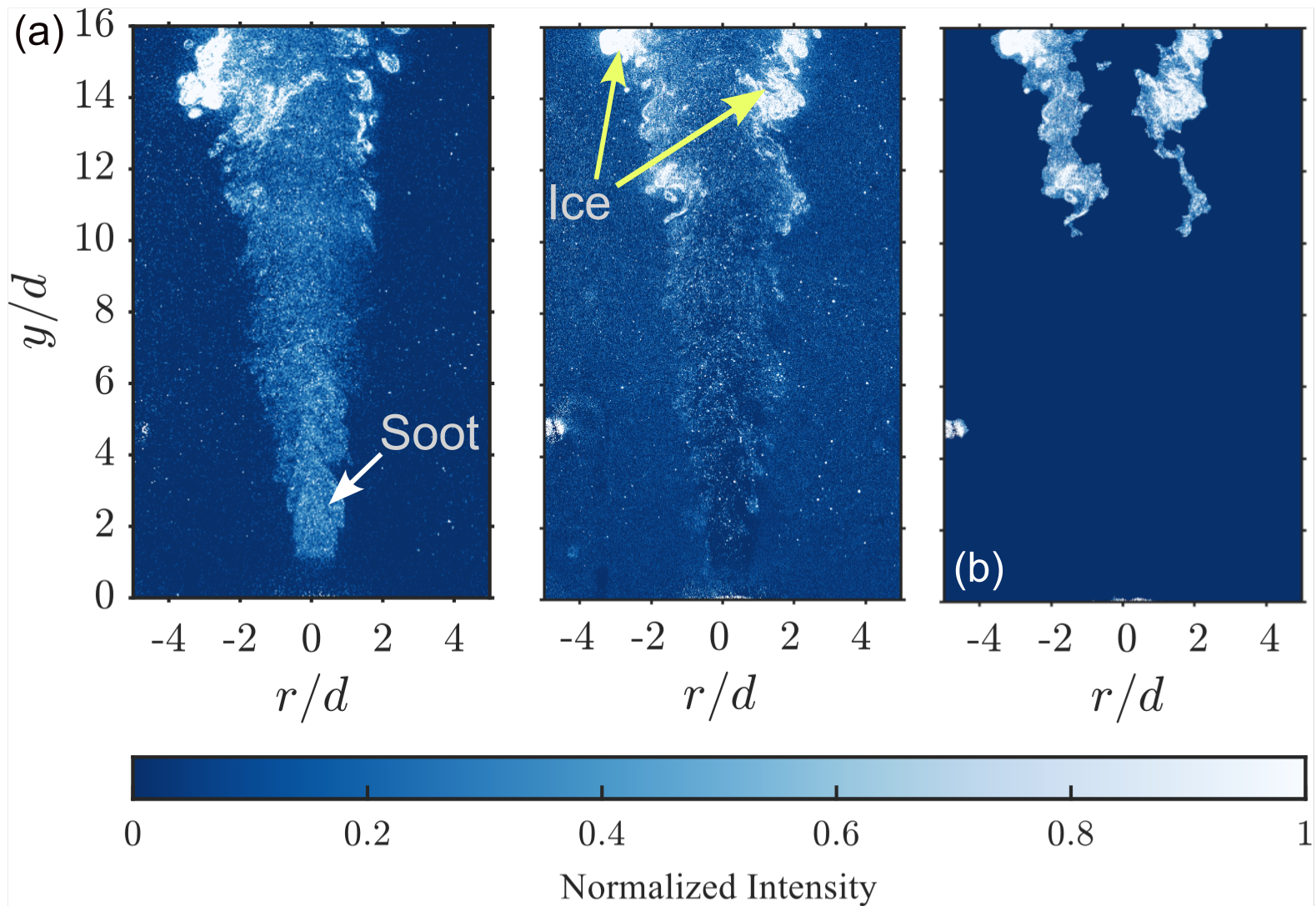}
	\caption{\footnotesize (a) Instantaneous scattering at target conditions revealing soot along the jet centerline and copious ice formation along the shear layers from exhaust of $\phi=0.161$ \ce{C2H4} flame. (b) Isolated scattering by ice particles after thresholding and removing soot scattering.}
	\label{fig:scatteringInst}
    \vspace*{-15pt}
\end{figure}

Background images at all polarization vectors are recorded with steady state conditions of cold ambient and a heated central jet, but without the presence of soot particles or exhaust composition. Measurements are then performed under soot-exhaust conditions by acquiring 500 scattering images containing soot and ice particles at multiple polarization vectors. A photo-diode is used to correct shot-to-shot laser power variation. All images are background-subtracted, denoised using a median filter, and time-averaged to obtain average scattering statistics (processing algorithm detailed in \S3 of SI). Instantaneous raw images are shown in Fig. \ref{fig:scatteringInst}a. Scattering from soot particles is also measured and corrected for under identical flow conditions and ambient-temperature. Mean scattering intensity solely due to ice particles $I^{\rt{sca}}_{\rt{ice}}(\bm{x})$ is then obtained by subtracting the mean contributions of soot scattering from the combined scattering intensity. For instantaneous images like Fig. \ref{fig:scatteringInst}(a,b), ice scattering can be distinguished from soot using an intensity based thresholding method. Since soot particles are of sub-micrometer size, ice scattering can be much more intense \cite{bohren2008scattering} and can easily be distinguished from soot scattering. 

Ice formation is analyzed using polarization-based measurements of the Stokes' vectors \cite{schaefer2007measuring}. In the presence of scatterers, the incident and scattered Stokes vectors are related to each other through the scattering matrix ($\bmss{M}$) as: $\bmss{I}^{\rt{sca}}=\bmss{M}\bmss{I}^{\rt{inc}}$, where the Stokes vector $\bmss{I}=[I\enspace Q\enspace U\enspace V]^\top$ contains components that can be used to decompose light into its myriad polarization states. The perpendicularly polarized incident light for our case is given by $\bmss{I}^\rt{inc}=I_{\rt{i}}[1\enspace-1\enspace 0\enspace0]^\top$. The polarization state of the scattered light can be determined from the intensity measured by varying the angles ($\theta_n$) between the quarter-wave plate and linear polarizer, and is given by \cite{schaefer2007measuring}:
\begin{align}
    I^{\rt{sca}}_{\rt{ice}}=\frac{1}{2}(A+B\sin2\theta_n+C\cos4\theta_n\notag\\
    +D\sin4\theta_n),\enspace n=1,\ldots,8,
\end{align}
where, $\theta_n$ is varied in steps of $\Delta\theta=\pi/8=22.5^\circ$ and the coefficients $A,B,C,D$ are determined by matching terms in the Fourier expansion: $I^\rt{sca}_\rt{ice}(\theta)=1/N_\theta \sum_{n}^{N_\theta}[I^\rt{sca}_\rt{ice}(\theta_n)\sin(n\theta_j)+I^\rt{sca}_\rt{ice}(\theta_n)\cos(n\theta_n)]$. The elements of $\bmss{I}^\rt{sca}$ are then determined as: $I_{\rt{s}}=A-C,\enspace Q_{\rt{s}}=2C, \enspace U_{\rt{s}}=2D, \enspace V_{\rt{s}}=B$. The vertical and horizontal polarization components arising from vertically polarized incident light is obtained as: $I^\rt{sca}_{\rt{vv}}=0.5(\sqrt{Q^2_{\rt{s}}+U^2_{\rt{s}}+V^2_{\rt{s}}}-Q_{\rt{s}})$ and $I^\rt{sca}_{\rt{hv}}=0.5(\sqrt{Q_{\rt{s}}^2+U_{\rt{s}}^2+V_{\rt{s}}^2}+Q_{\rt{s}})$, respectively. Vertically polarized light incident on perfectly spherical Rayleigh scatterers is expected to scatter light while retaining its polarization, such that $\bmss{I}^\rt{sca}=I_{\rt{s}}[1\enspace-1\enspace 0\enspace0]^\top$. However, small changes ($\sim 5\%$) in the polarization are expected from real soot aggregates \cite{sorensen2018rdg} and multiple scattering contributions. Larger changes are, however, expected from asymmetrically-shaped scatterers and optically dense scattering media. In fact, depolarization of back-scattered Lidar is a commonly used method for distinguishing water droplets from ice in mixed-phase clouds \cite{sassen2001midlatitude, urbanek2018high} and distributions of various ice habits in cirrus clouds \cite{noel2005study, okamoto2019interpretation}. Thus, we define an asymmetry factor
\begin{equation}    \rho_\rt{i}=\sqrt{U_{\rt{s}}^2+V_{\rt{s}}^2}/\sqrt{Q_{\rt{s}}^2+U_{\rt{s}}^2+V_{\rt{s}}^2},
\end{equation} 
which relates the induced circular polarization by ice crystals due to their non-spherical shape from initially linearly polarized light.

\subsection{Favre averaged Navier-Stokes simulations with soot/ice aerosol model}\addvspace{5pt}
Further contextual information is extracted from numerical simulations and used to complement experimental results. Favre-time-averaged Navier Stokes (FANS) equation with a $k-\omega$ turbulence closure is solved for the present geometry along with semi-empirical two equation model for monodisperse soot and ice particulates with inputs from experimental results. Some salient details are briefly introduced here and the numerical setup for the geometry including flow validation is included in \S3 of SI. A more comprehensive discussion has been presented elsewhere \cite{chelem2025contrail, jahncke2026fans}. 

In addition to the flow equations, the transport equation of the Favre-averaged mass fraction $\tilde{Y}_p$ and number density per unit mass ($\tilde{N}_p$) of the dispersed phase (soot and ice) are solved simultaneously (Eq. 4,5 in 
\S3 of SI). When identified as soot, the dispersed phase is assumed to be non-reactive, monodisperse and spherical with the same diameter as the mode of the distribution described in Fig. \ref{fig:mobdia}. For ice, the dispersed phase remains monodisperse as ice continues to accumulate mass on soot particles and grow in spherical habit. The monodisperse treatment presents a limitation of the model as it disregards the size-dependent activation \cite{testa2024soot, testa2024simulated} and growth of ice crystals \cite{gao2022enhanced}. 

\begin{figure}[t]
	\centering
	\includegraphics[width=\linewidth]{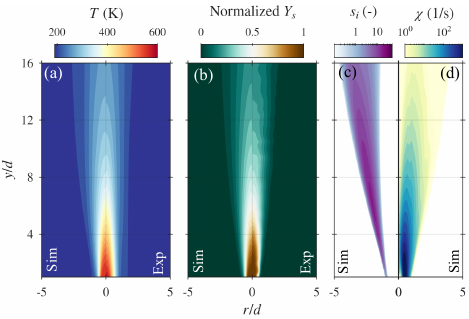}
	\caption{\footnotesize (a) Temperature field which is maintained constant for all the experimental conditions. (b) Normalized $N_\rt{s}$ from exhausts of \ce{C2H4} flame at $\phi=0.161$ obtained from simulations by setting $\overline{\dot{\omega}}_\rt{ice}=0$ in Eq. \eqref{omegaIce01} and from scattering measurements at room temperature. (c), (d) Supersaturation over ice and scalar dissipation rate in \ce{C2H4} case at $\phi=0.124$, obtained from simulations by setting $\overline{\dot{\omega}}_\rt{ice}=0$ in Eq. \eqref{omegaIce01}}
	\label{fig:TempProfile}
    \vspace*{-15pt}
\end{figure}

A one-step kinetic model for heterogeneous ice nucleation that depends on local flow properties and saturation conditions is considered here \cite{khou2015spatial, chelem2025contrail}. Higher-order interactions of dispersed phase with turbulence is neglected. All soot particles are assumed to be activated and serve as ice-nucleating particles (INPs). This assumption is strictly not true given that freshly generated soot particles are poor INPs \cite{seisel2005water, testa2024simulated}. However, exceedingly high level of super-saturation over ice ($s_i>>10$) and cooling rates in the shear layers (in Figs. \ref{fig:TempProfile}c, \ref{fig:TempRadialProfile}) counter the hydrophobic nature of soot and is expected to lead to a more rapid and significantly higher levels of activation, particularly for the cases studied here. The rate of water vapor condensation is determined through a modified Fick's law for particles whose
radius $r_{p}$ are on the order of the mean free path. Under such considerations, the net rate of mass transfer to ice particles is given by \cite{chelem2025contrail, jahncke2026fans}:
\begin{align}
\overline{\dot{\omega}}_\rt{ice} = &\frac{4\pi \tilde{N}_\rt{p} r_\rt{p} D_\rt{vap} M_{\ce{H2O}}}{R\overline{T}}(\overline{p}_\rt{vap}- \overline{p}_\rt{vap}^\rt{sat,i})\notag\\
&G(r_\rt{p})\Pi(\overline{p}_\rt{vap}^\rt{sat,w}, r_\rt{p}),
\label{omegaIce01}
\end{align}
with diffusion coefficient of water vapor $D_\rt{vap}$, molecular weight of water $M_{\ce{H2O}}$, specific gas constant $R$, vapor partial pressure $\bar{p}_\rt{vap}$, saturation vapor pressure above liquid water $\bar{p}_\rt{vap}^\rt{sat,w}$, and ice $\bar{p}_\rt{vap}^\rt{sat,i}$. The factor $G(r_\rt{p})$ depends on the Knudsen number accounting for the transition from kinetic to diffusion regime as $r_\rt{p}$ changes. The thermodynamic condition for condensation is given by:
\begin{align}
\Pi=
\begin{cases}
0 \enspace\rt{if}\enspace \bar{p}_\rt{vap}\leq \overline{p}_\rt{vap}^\rt{sat,w} \& \enspace r_\rt{p}=r_\rt{s},\\
1 \enspace\rt{if}\enspace \bar{p}_\rt{vap}> \overline{p}_\rt{vap}^\rt{sat,w} \| \enspace r_\rt{p}>r_\rt{s},  
\end{cases}  
\end{align}
where $r_\rt{s}$ is the soot diameter that grows to $r_\rt{p}$ following ice nucleation.

A finite volume solver is used on a 2D axisymmetric mesh using a second-order accurate finite-volume spatial discretization scheme with anisotropic block-based adaptive mesh refinement and iterative quasi-Newton method to attain rapid convergence for the steady flow problem \cite{charest2010computational, charest2011experimental}. The mesh for the present geometry comprises a total of 1,830 grid blocks, with 4 $\times$ 8 cells per block, resulting in a total of 58,560 computational cells. The $k-\omega$ turbulence model has been systematically validated against benchmark numerical simulations \cite{xing2022quadrature, xing2022use, FUN3D, CFL3D_v67, chelem2025contrail} while the turbulent mixing have been validated against experimental datasets \cite{chelem2025contrail, guitton2007measurements}, and is adopted here for the current geometry. The exhaust composition at the central nozzle inlet is calculated based on the complete combustion of a given pair of fuel and global $\phi$, while the $N_\rt{s}$ and $d_\rt{m}$ are supplied from SMPS measurements (Fig. \ref{fig:mobdia}). Thermal boundary condition at the nozzle inlet and walls along with flow rate and pressure conditions are supplied from measured data. The measured temperature profile is used to validate the converged flow solutions (see Figure S.7 in SI). 

\subsection{Mie scattering solution}\addvspace{5pt}
The inversion of measured scattering data to ice number concentrations $\tilde{N}_p(\bm{x})$ and particle diameters $d_p(\bm{x})$ are currently unfeasible. So, we utilize Mie scattering calculations to convert numerically simulated data to scattering cross sections and offer comparisons with experimental measurements. Specifically, we calculate $I^\rt{sca}_\rt{vv}(90^\circ)$ using the MiePython library \cite{prahl2024miepython, bohren2008absorption}. Note that $I^\rt{sca}_\rt{hv}$ remains identically zero at $\theta^\rt{sca}=90^\circ$ as spherical ice particles would not induce horizontal polarization for incident light which is vertically polarized \cite{bohren2008absorption}. Deviations of scattered intensities from the vertical polarization are quantified as the asymmetry factor, $\rho_\rt{i}$. In the absence of a polydisperse treatment for particle size, and to avoid the resultant fluctuations in the Mie theory results for $x_p = \pi d_p/\lambda \sim 1$, we averaged the intensity over a log-normal distribution of particle sizes centered around the simulated value of $d_p(\bm{x})$ and a standard deviation of $0.3$ at each location \cite{karcher2015microphysical}. 
\addvspace{-5pt}
\section{Results\label{subsec:results}} \vspace{-5pt}

\subsection{Jet mixing, inception and contrail microphysics\label{subsec:aerodynamics}}\addvspace{5pt}

Contrail formation is initiated by the scalar mixing of temperature ($T$), water vapor ($x_{\ce{H2O}}^\rt{v}$) and soot particles by the turbulent jet flow. The mean flow jet profile of these variables is shown in Figure \ref{fig:TempProfile}. With sufficiently resolved boundary conditions as in current experiments, the numerical simulations capture the centerline and radial decay of $T$ and soot mass fraction $Y_s$ quite well. Some difference is noted in the profile for $Y_s$, which can be observed to decay faster in experiments. This is possibly due to the fact that $Y_s$ is inferred from scattering measurements at room temperatures so as to avoid ice nucleation. Overall, both $T$ and $Y_s$ follow a similar decay and spreading profile.  
This is expected since the Lewis number $Le=\alpha_t/\mathcal{D}_t$ associated with the thermal ($\alpha_t$) and scalar ($\mathcal{D}_t$) diffusivity seldom deviates from unity, and so the spread of scalar and soot particles are quite similar \cite{karcher1994transport, paoli2004contrail}.  

Although not measured, it is assumed that $x_{\ce{H2O}}^\rt{v}$ has a similar spreading profiles and decay. As a result, $x_{\ce{H2O}}^\rt{v}$ and $T$ decay proportionally while the saturation pressures $p_\rt{vap}$ drops exponentially, resulting in a supersaturated zone before the the scalar decays to the ambient sub-saturated flow. The resultant supersaturation profiles over ice, $s_i=p_\rt{vap}/p_\rt{vap}^{\ce{sat, i}}-1$, is shown in Fig. \ref{fig:TempProfile}(c) for \ce{C2H4} flame at $\phi=0.124$ simulated by keeping $\overline{\dot{\omega}}_\rt{ice}=0$ in Eq. \eqref{omegaIce01}. Even for the lowest water containing exhaust considered in this study (\ce{C2H4} flame at $\phi=0.124$), the supersaturation is noted to  achieve a maximum value of $\sim 50$. For a typical engine exhaust and at these high supersaturation values, nearly all soot particles get activated and subsequently freeze, despite their low hygroscopicity \cite{karcher2015microphysical}. Recent measurements with SAF, however, have noted that such fuels can reduce particle activation by $14\%$ to $52\%$\cite{leclercq2022emission, voigt2021cleaner, markl2024powering}. Since we lack the quantification of physicochemical characteristics on soot particles, the assumption of complete soot activation reflects the limiting case for which the model is expected to over‑predict ice formation.

\begin{figure}[t]
	\centering	\includegraphics[width=0.48\textwidth]{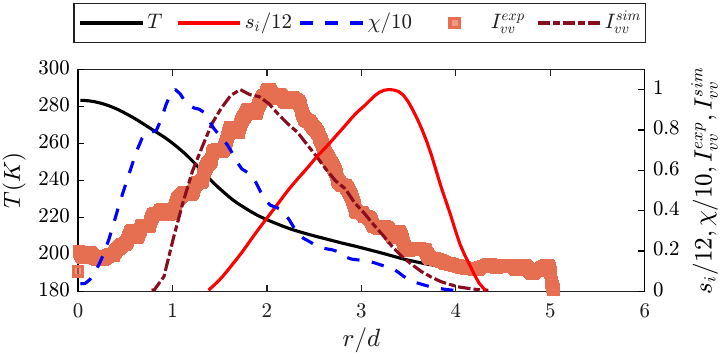}
	\caption{\footnotesize Radial variation of $T$, supersaturation over ice $s_\rt{i}$ and scalar dissipation rate $\chi$ at $y/d=14$ for $\phi=0.124$ \ce{C2H4} flame. Ice formation is driven by the inflection in the measured scalar profiles.}
	\label{fig:TempRadialProfile}
    \vspace*{-15pt}
\end{figure}
The turbulent scalar mixing governs not only the cooling rates, and consequently, the ice nucleation timescales \cite{karcher2015microphysical, barahona2012ice}, but also mixing and growth timescales. Consequently, we show the diffusivity governed decay rate of scalar variance using scalar dissipation rates $\chi=\overline{2\mathcal{D}_t(\nabla T)^2}$ in Fig. \ref{fig:TempProfile}(d), the inverse of which governs most relevant timescales. The scalar dissipation is maximized near the boundary of jet and ambient flow ($\chi \sim \SI{1000}{s^{-1}}$) and decreases with decreasing gradients downstream to $\chi \sim \SI{1}{s^{-1}}$. Naturally, higher cooling rates and supersaturation ensure that most particles activate and nucleate close to the jet exhaust. Further downstream, while nucleation rates are reduced, the timescales available for growth are higher. The consequences of scalar mixing is reflected in the instantaneous snapshot of scattering from soot and inchoate ice particles (Fig. \ref{fig:scatteringInst}). Significant amount of ice formation is observed to develop along the shear layers where $T$, $p_\rt{vap}$ and $N_\rt{s}$ is lowered while the water vapor saturation $s_i=p_\rt{vap}/p_\rt{vap}^\rt{sat, i}(T)$ over ice and water increases. This is depicted in Fig. \ref{fig:TempRadialProfile} which shows the radial profiles of $T(r)$, $s_\rt{i}$ and $I_\rt{vv}$ from simulation and experiments for \ce{C2H4} flame at $\phi=0.124$. Region of intense mixing is marked by the peak in $\chi$, which closely corresponds to a region where flow becomes supersaturated ($s_i>0$), resulting in rapid ice formation and growth as captured by $I_\rt{vv}$. Ice formation peaks at $T\sim220$ K after which scattering rapidly decays due to a decay in soot particles available for nucleation. Finally, we note the difference in $I_\rt{vv}^
\rt{exp}$ and $I_\rt{vv}^\rt{sim}$ which possibly arises from a number of simplifying assumptions in our modeling.  
\begin{figure}
    \centering
    \includegraphics[width=\linewidth]{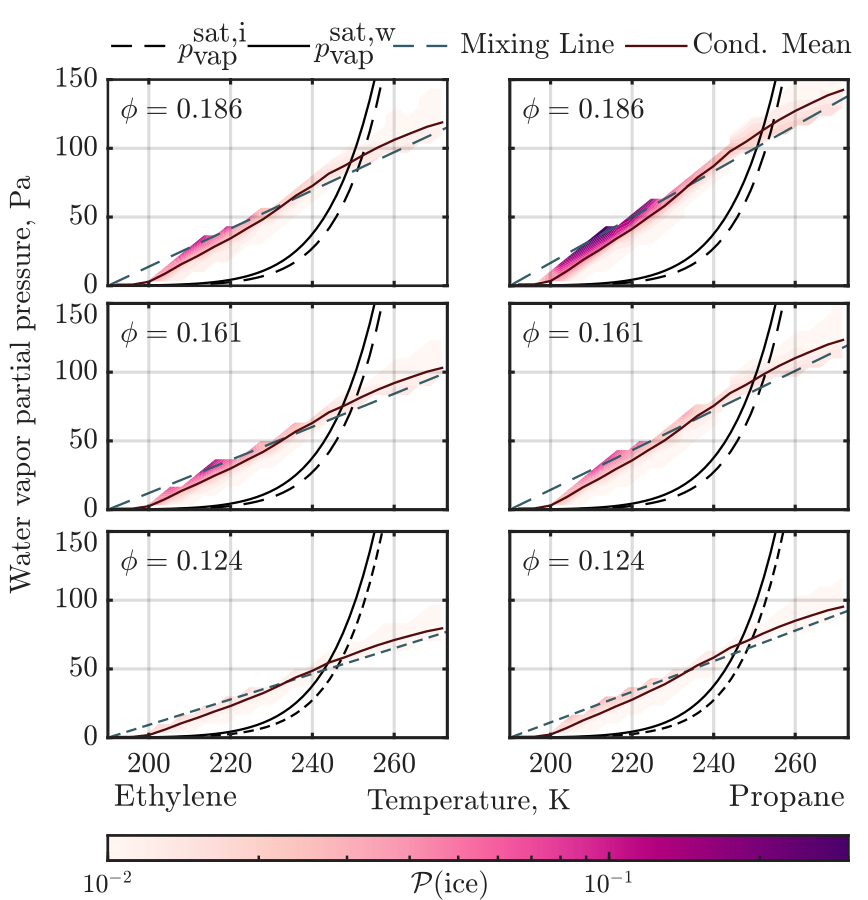}
    \caption{Probability of ice formation obtained from instantaneous ice scattering for different experimental conditioned on $p_\rt{vap}$ and $T$. An increased probability of ice formation is observed with increasing $N_\rt{s}$ across all values of $\phi$ and with increasing $x^\rt{v}_{\ce{H2O}}$ emissions from \ce{C3H8}. }
    \label{fig:SAC_prob}
    \vspace{-17pt}
\end{figure}

In order to gain further insights into the mixing process, we transform the intensity of instantaneous ice scattering into probability of finding ice $\mathcal{P}(\rt{ice})$ at a given location conditioned on the $\bar{T}$ and $\bar{p}_\rt{vap}$. We calculate $p_\rt{vap}=p_\infty x^\rt{v}_{\ce{H2O}}$, where it is assumed that $x^\rt{v}_{\ce{H2O}}$ scales in the same way as the normalized $Y_s$ (Fig. \ref{fig:TempProfile}b). The conditioned probability is then expressed in the familiar $p_\rt{vap} - T$ coordinates in Fig. \ref{fig:SAC_prob} for aiding comparison with the Schmidt-Appleman criteria \cite{schmidt1941entstehung, appleman1953formation,paoli2016contrail}. The ice ($p^{\mathrm{ice/sat}} _{\mathrm{vap}}$) and liquid water ($p^{\mathrm{liq/sat}} _{\mathrm{vap}}$) saturation curves are also shown. Regions above the ($p^{\mathrm{liq/sat}} _{\mathrm{vap}}$) represent the supersaturated regions where ice nucleation occurs. Since mixing happens along an isobaric line, the mixing process typically follows a straight line in the $p_\rt{v} - T$ coordinates, with exhaust conditions as the starting point and ambient conditions as the final. Naturally, the slope increases with increasing $x^\rt{v}_{\ce{H2O}}$ in the exhaust and is consequently higher for propane flames (Fig. \ref{fig:SAC_prob}d-f) at similar $\phi$ values. As the slope increases, the maximum supersaturation increases, and so does $\mathcal{P}(\rt{ice})$. Further, the probability of finding ice among the cases is found to be proportional to their $x^\rt{v}_{\ce{H2O}}$ and $N_\rt{s}$. Thus, higher supersaturation environments allow smaller vortical structures with shorter timescales to complete nucleation and growth, thereby increasing the probability of finding ice in those locations.   
\begin{figure*}[t]
	\centering
\includegraphics[width=\textwidth]{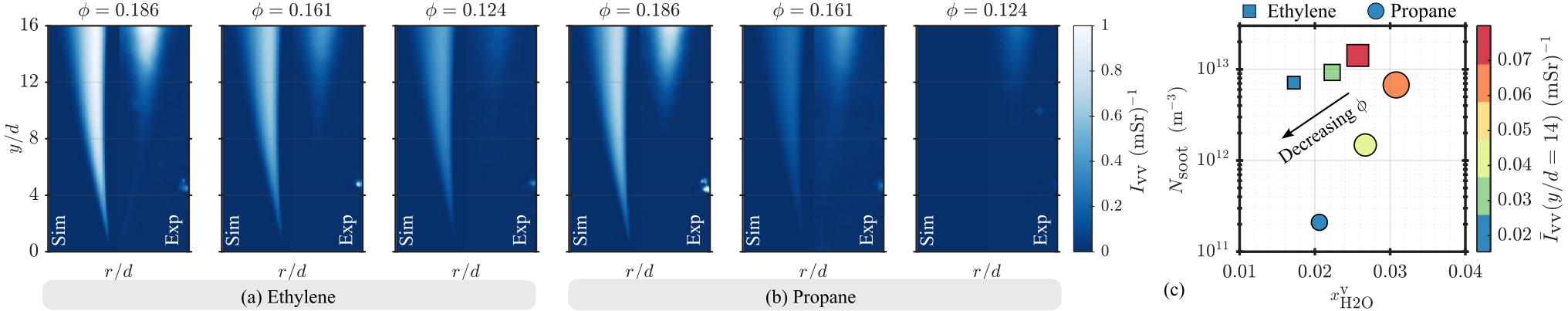}
	\caption{\footnotesize (a-b) $I_{\rt{vv}}$ evaluated from combining simulations and Mie theory (left panel) and observed experimentally through Stokes polarimetry (right panel). Significant values of $I_{\rt{vv}}$ confirm ice nucleation for the different cases. (c) Radially averaged scattering intensity at $y/d=14$ plotted as a function of $N_{\text{s}}$ and $x_{\ce{H2O}}^\rt{v}$.}
	\label{fig:scatteringMean}
    \vspace*{-15pt}
\end{figure*}
\addvspace{10pt}
\subsection{Mean scattering from ice particles in numerical simulations and experiments\label{subsec-meanscattering}}\addvspace{5pt}


Figure \ref{fig:scatteringMean} shows the mean elastic scattering intensity $\bar{I}_\rt{vv}$ by ice particles from experiments (right) and their comparisons with numerical simulations evaluated through Mie scattering solution (left). We note that the scattering intensity decreases with decreasing equivalence ratio for exhausts from ethylene and propane flames, despite increments in soot diameter and fractal dimension (Table \ref{tab:exp_cond}). This is because water vapor content and $N_\rt{s}$ simultaneously drop with decreasing $\phi$, thereby overcoming the marginal effect of poorer INPs. For \ce{C2H4} exhausts, $\bar{d}_\rt{m}$ remains identical, so the change in $I_\rt{vv}$ scales with decreasing $N_\rt{s}$ at globally leaner exhausts. On the other hand, for $\ce{C3H8}$ exhausts, $\phi=0.161$ has lower $\bar{d}_\rt{m}$ and $N_\rt{s}$ than $\phi=0.186$. So, the decrease in intensity is due to a combination of smaller ice particles formed at an almost lower order of magnitude concentration. Across the two fuels, propane flames generate lower $N_\rt{s}$ when compared to ethylene flames, but contain higher water vapor content in the exhaust (Table \ref{tab:exp_cond}). Typically, this results in lower scattering coefficients, highlighting the benefits of lower PM content. The benefits of lower PM content, for typically higher sooting flames, in avoiding contrail scattering are also noted in the comparison of \ce{C2H4} flame with $\phi=0.186$ and \ce{C3H8} flame with $\phi=0.161$. Each of these flames have a similar $x^\rt{v}_{\ce{H2O}}$, but the lower sooting propane flame results in much smaller scattering cross sections.  

A confounding behavior is however seen for $\ce{C3H8}$ flame exhausts at $\phi=0.124$ (Fig. \ref{fig:scatteringMean}b). For this case, the nvPM number concentration remained significantly lower than other cases, and yet led to detectable scattering due to ice. This anomalous increase in $I^\rt{sca}_\rt{vv}$ might be attributed to the presence of ultra-fine volatile particulate matter in the exhaust of leaner \ce{C3H8} MISG flames (as discussed in \S\ref{subsec:flow-particle},  \cite{moallemi2019characterization, vernocchi2022characterization}). The vPM content is not measured by the SMPS, but would have the potential to activate into ice and result in the observed scattering. 
Such a trend has recently been proposed where substantial ice formation was recorded from cleaner burning aircraft engine solely due to the presence of volatile content in the exhaust \cite{karcher2018formation, voigt2025substantial}.

Interestingly, we note that the simulations capture the overall trends in the scaling of scattering cross-section with varying $\phi$ for the two fuels. However, the model over-predicts the amount of ice formation for most cases. In contrast, for $\phi=0.161$ and $\phi=0.124$ $\ce{C3H8}$ cases, the model under predicts the scattering cross-section, potentially due to the presence of vPM in the exhaust. As the present model does not account for vPM, the Mie scattering cross-section calculated based on just the nvPM of $N_\rt{s}=0.21$ and $1.48\times 10^{12}$ (m$^{-3}$), for the cases respectively, would be negligible, as is reflected in the comparison. Further, the activation and growth of ice particles also takes place much closer to the nozzle in contrast to the experimentally observed trends. Thus, the present model is a limiting case which provides useful upper bounds of ice formation. The results also provide context and validation for the two-equation model which is used quite widely in the literature \cite{khou2015spatial,  cantin2022eulerian, vancassel2014numerical, afkari2026evaluating}. Naturally, the present model can be improved by the inclusion of more physical activation functions \cite{knopf2023atmospheric, marcolli2020soot}, accounting for poly-dispersed  nature of aerosols \cite{xing2022quadrature, xing2022use}, consideration for volatiles along with experimental quantification of surface chemistry parameters such as soot solubility, volatile content and organic coatings \cite{gao2022dependence}.

To see the dependence of scattering-cross section on $N_\rt{s}$ and $x_{\ce{H2O}}^\rt{v}$ more clearly, Fig. \ref{fig:scatteringMean}(c) presents the radially averaged scattering intensity at a downstream location of $z/d=14$ for the six cases in a two-parameter plot. It can be seen that for a given fuel, the increase in water vapor and soot causes an increase in scattering intensities, while comparison among different fuels highlight its higher sensitivity to water vapor concentrations.

\begin{figure}[h!]
    \centering
    \includegraphics[width=\linewidth]{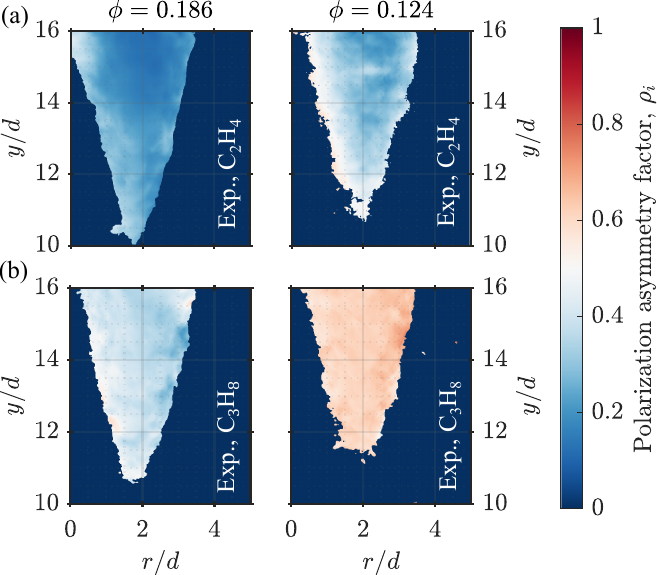}
    \caption{Asymmetry factor $\rho_\rt{i}$ quantifying the induced circular polarization by ice particles, implying their non-spherical nature. Note only the zone around ice cloud ($y\geq10d$) is visualized.}
    \label{fig:asym}
    \vspace{-10pt}
\end{figure}

One final question we have not addressed yet: what are the properties of ice particles? To this end, we evaluate the asymmetry factor, $\rho_\rt{i}$, of the contrail cloud, which quantifies the degree of induced circular polarization from incident vertical polarization. The asymmetry factor then quantifies the non-sphericity of ice crystals within the jet plume. This is shown for a select few cases in Fig. \ref{fig:asym}. With decreasing $\phi$, $\rho_\rt{i}$ decreases for both ethylene and propane flames, but is noted to be consistently higher for propane flames. Since propane flames result in comparable or higher ice-nucleation probabilities but lower particulate concentrations, the larger supersaturation must sink to fewer particles (nvPM or otherwise), possibly resulting in larger, more irregularly shaped particles (Fig. \ref{fig:asym}b). While simulations support these results by predicting much larger particles for propane flames than ethylene flames (not shown here), it does not capture irregular shapes of ice habits as it only assumes spherical particles. Discriminating various ice habits will be explored in the future.

\section{Conclusions\label{sec:conclusions}} 
To summarize, we have developed a novel, lab-scale facility for estimating the impact of aviation on atmosphere due to the formation of contrail cirrus. By combining sensitively controlled experiments, multi-dimensional optical diagnostics and well-constrained numerical modeling, we explored how fuel and soot content, influenced by changing $\phi$ in \ce{C2H4} and \ce{C3H8} flames, might play a role in determining the contrail-forming propensity. 

Mie scattering images of the contrails revealed the turbulent structures that initiate the mixing process, leading to the microphysical activation and growth of soot aerosol into ice particles within single vortical structures. Instantaneous conditional probabilities of ice formation along the shear layer was mapped on to the Schmidt-Appleman diagram. This probability of finding ice is noted to depend on local supersaturation and increases with increasing exhaust water vapor content. Further, experimental observations were coupled with Favre-averaged Navier-Stokes (FANS) along with transport equation for the dispersed phase and empirical diffusional model of ice nucleation. Simulated data was transformed into scattering cross-sections by assuming spherical ice particles, and was shown to closely follow experimentally observed trends of decreased scattering intensity with reduction in $\phi$ and $N_\rt{s}$ across both the fuel cases. Finally, depolarization measurements utilizing Stokes polarimetry was used to ascertain aspheric nature of ice particles that form in the shear layers. 

Thus, our study presents an unprecedented window into the fascinating phenomena of contrails and opens up the possibility of reliably estimating the impact of future fuels on their contrail-forming potential.



\vspace{-2pt}
\acknowledgement{CRediT authorship contribution statement} 

{\bf Amitesh Roy \& Rajat Sawanni}: Contributed equally to conceptualization, methodology, investigation, data curation, validation, writing - original draft. {\bf Yash T. Rajan, Isaac Jahncke \&  Taye Taddesse}: Methodology, investigation, data curation, validation, writing - review and editing. {\bf Clinton P. T. Groth, Swetaprovo Chaudhuri, \"Omer L. G\"ulder}: Conceptualization, project administration, funding acquisition, methodology, resources, supervision, writing - review and editing.

\acknowledgement{Declaration of competing interest} 
The authors declare that they have no known competing financial interests or personal relationships that could have appeared to influence the work reported in this paper.
\vspace{-2pt}
\acknowledgement{Acknowledgments} 
The authors acknowledge support from NSERC Alliance Missions (ALLRP 570573-2021) and NSERC Discovery Grants, and helpful discussions with Russell Stratton and John Hu from Pratt \& Whitney Canada. AR, RS and YTR are grateful to all the labmates who helped establish the contrail facility.

\footnotesize
\baselineskip 9pt

\thispagestyle{empty}
\bibliographystyle{proci}
\bibliography{ref}

\ifarXiv
  \includepdf[pages=-]{\supplementfilename}
\fi



\newpage

\small
\baselineskip 10pt


\end{document}